\documentclass[11pt]{article}
\usepackage[normalem]{ulem}
\pdfoutput=1 % if your are submitting a pdflatex (i.e. if you have

\usepackage{jheppub} % for details on the use of the package, please
\usepackage{url}
\usepackage{caption}
\usepackage{subcaption}
\usepackage{extarrows}

\usepackage[latin9]{inputenc}
\usepackage{float}
\usepackage{amsmath}
\usepackage{amssymb}
\usepackage{graphicx}
\usepackage{esint}
\usepackage{hyperref}
\usepackage{comment}
\usepackage{color}
\usepackage{microtype}
\usepackage{cleveref}
\usepackage{breakurl}
\usepackage{bbm}
\newcommand{\be}{\begin{equation}}
\newcommand{\ee}{\end{equation}}
\newcommand{\ben}{\begin{displaymath}}
\newcommand{\een}{\end{displaymath}}
\newcommand{\bea}{\begin{eqnarray}}
\newcommand{\eea}{\end{eqnarray}}
\def\K{K{\"a}hler }
   \newcommand{\rf}[1]{(\ref{#1})}
\newcommand{\vp}{\varphi}

\def\be{\begin{equation}}
\def\ee{\end{equation}}
\def\bea{\begin{eqnarray}}
\def\eea{\end{eqnarray}}
\def\ba{\begin{array}}
\def\ea{\end{array}}
\def\bit{\begin{itemize}}
\def\eit{\end{itemize}}

\def\a{\alpha}

\def\vp{\varphi}
\def\m{\mu}

 \makeatletter

\allowdisplaybreaks

\makeatother

\makeatletter
\DeclareRobustCommand{\rcite}[1]{%
  \rcite@aux#1,\@nil{#1}%
}
\def\rcite@aux#1,#2\@nil#3{%
  \if\relax#2\relax
    Ref.~\cite{#3}%
  \else
    Refs.~\cite{#3}%
  \fi
}
\makeatother

\hypersetup{
    colorlinks = true,
    citecolor = {blue},
    linkcolor = {blue},
    urlcolor = {blue},
}

 \title{\rm { \huge  \bf   Polynomial   {\boldmath $\a$}-attractors   }}

\author{Renata Kallosh \ and }
\author{Andrei Linde}

\affiliation{Stanford Institute for Theoretical Physics and Department of Physics,\\ Stanford University, Stanford, CA 94305, USA}
\emailAdd{kallosh@stanford.edu}
\emailAdd{alinde@stanford.edu}

\abstract{Inflationary $\alpha$-attractor models can be naturally implemented in supergravity with  hyperbolic geometry. They have stable predictions for observables, such as  $n_s=1-{2/ N_{e}} $, assuming that the potential in terms of the original geometric variables, as well as its  derivatives, are not singular at the boundary of the hyperbolic disk, or half-plane. In these models, the potential in the canonically normalized inflaton field $\varphi$ has a plateau, which is approached exponentially fast at large $\varphi$. We  call them {\it exponential $\alpha$-attractors}. We present a closely related class of models, where the potential is not singular, but its  derivative is singular at the boundary. The resulting inflaton potential is also a plateau potential, but it approaches the plateau polynomially. We  call them {\it polynomial $\alpha$-attractors}. Predictions of these two families of attractors completely cover the sweet spot of the Planck/BICEP/Keck data.   The exponential ones are on the left, the polynomial are on the right.}

\begin{document}

\maketitle

% \tableofcontents{}
%\newpage

\section{Introduction}

\parskip 3pt

 The recent data release from BICEP/Keck   \cite{BICEPKeck:2021gln} (see also \cite{Tristram:2021tvh}) considerably strengthened  bounds on the tensor to scalar ratio $r$. Their results ruled out several popular inflationary models, such as the models with monomial potentials, the original version of the natural inflation scenario, and the models with the Coleman-Weinberg potentials  previously used in new inflation.  On the other hand, there are several well-known models which fit all available data  \cite{Kallosh:2021mnu}. Some of these models, such as the Starobinsky model \cite{Starobinsky:1980te},  the GL model \cite{Goncharov:1983mw,Goncharov:1984jlb,Linde:2014hfa},  the Higgs inflation \cite{Futamase:1987ua,Salopek:1988qh,Bezrukov:2007ep}, have been  proposed long ago.  Recently,  these models have been incorporated in the context of the cosmological $\alpha$-attractors \cite{Kallosh:2013hoa,Ferrara:2013rsa,Kallosh:2013yoa,Galante:2014ifa,Kallosh:2015zsa,Kallosh:2019eeu,Kallosh:2019hzo}.  
 
\begin{figure}[H]
\centering
\includegraphics[scale=0.09]{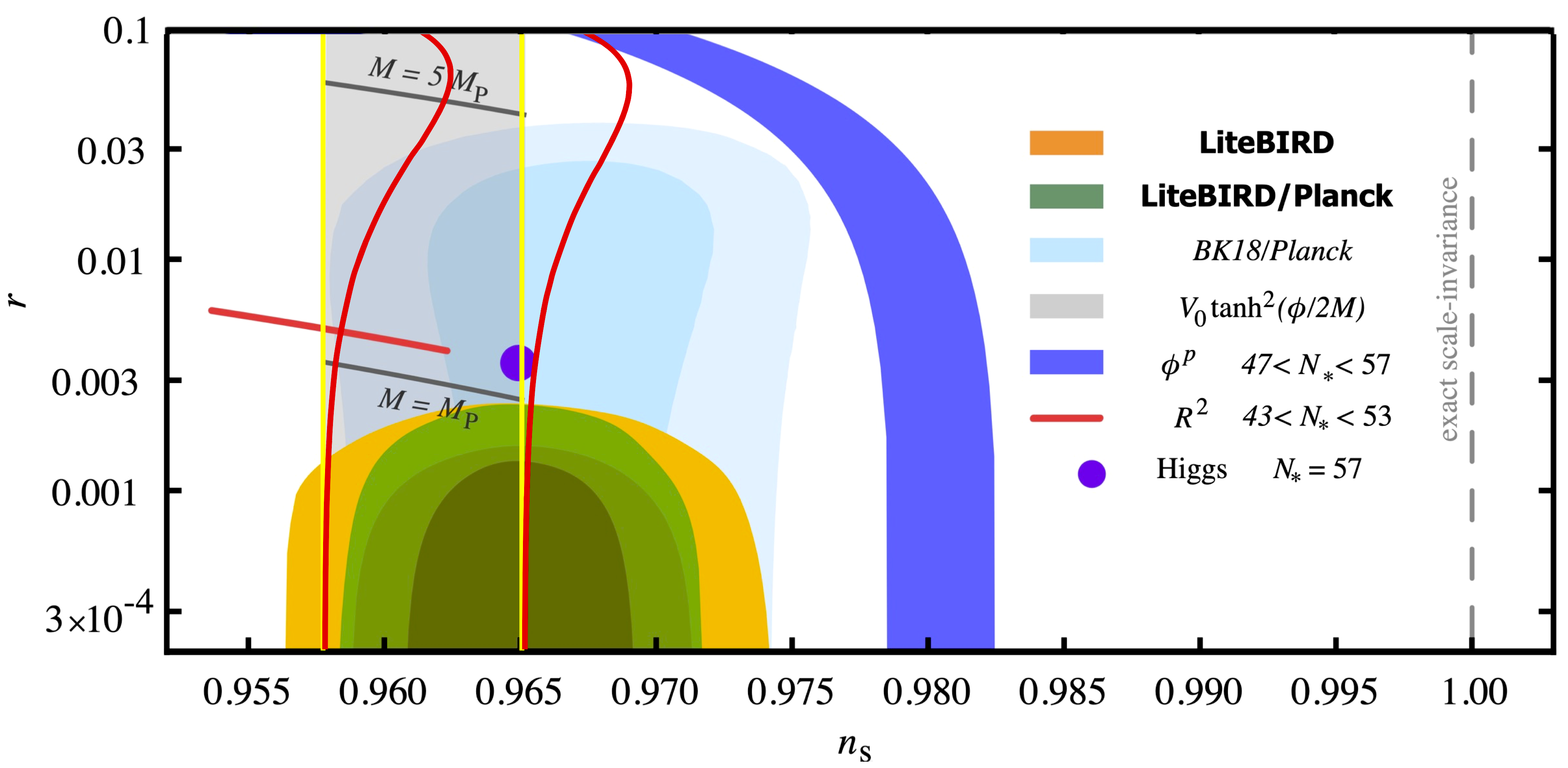} 
\vskip -5pt
\caption{\footnotesize The figure illustrating B-mode targets for LiteBIRD \cite{LiteBIRD:2022cnt}. The gray area shows the predictions of the simplest $\alpha$-attractor T-models with the potential $V \sim \tanh^{2} {\varphi\over \sqrt{6\alpha}}$. It is surrounded by two yellow lines corresponding to the number of e-foldings $N_{e} = 47, 57$. We added two red lines for $N_{e} = 47, 57$ surrounding predictions of  E-models of $\alpha$-attractors with the potential $V\sim \big (1-e^{-\sqrt{{2\over 3\alpha}}\varphi}\big )^{2}$  \cite{Kallosh:2013yoa}.  Predictions of these  models  cover the left half of the blue area favored by  \cite{BICEPKeck:2021gln}. However, this LiteBIRD figure from  \cite{LiteBIRD:2022cnt} does not contain any targets corresponding to the right part of the blue area.}
\label{LiteBIRD}
\end{figure}

 In Fig. \ref{LiteBIRD} we show the figure from the recent LiteBIRD collaboration paper ``Probing Cosmic Inflation with the
  LiteBIRD Cosmic Microwave Background Polarization Survey'' \cite{LiteBIRD:2022cnt}. As one can see, all B-mode targets in this figure are in the left side of the blue $n_s-r$ area favored by Planck/BICEP/Keck, and there are no inflationary model targets on the right hand side. The ones on the left include the gray band corresponding to the simplest T-model  $\alpha$-attractors with the potential   \cite{Kallosh:2013yoa}
  \be\label{T}
  V= V_0 \tanh^2 \big ({\varphi \over \sqrt{6\alpha} } \big). 
  \ee
  We added to this figure two red lines surrounding the band corresponding to E-models with the potential   \cite{Kallosh:2013yoa}
  \be\label{E}
  V\sim \big (1-e^{-\sqrt{{2\over 3\alpha}}\varphi}\big )^{2}
  \ee  
  In each of these two bands, for T- and E-models, there are seven specific targets corresponding to $3\alpha = 1,2,...,7$  \cite{Ferrara:2016fwe,Kallosh:2017ced,Gunaydin:2020ric,Kallosh:2021vcf}. These models are related to  Poincar\'e disks and inspired by string theory, M-theory and maximal supergravity. These 7 disks are presented in Fig. 2 in \cite{LiteBIRD:2022cnt}.

Kinetic terms in E-models are based on the $SL(2, \mathbb{R})$ symmetry, and T-models have  the $SU(1,1)$ symmetry. These symmetries of kinetic terms are slightly broken by the potentials; the main predictions of these models are determined by kinetic terms. A small difference between the predictions of E- and T-models before they reach the attractor point is due to a slightly different way of breaking of these symmetries by the potentials.   
 All of these models have plateau potentials  which  exponentially fast reach the plateau at large values of the inflaton field,
 \be \label{exp}
V = V_{0} (1-e^{-\vp/\mu}+...)  \ .
\ee 
We will call these models `exponential $\alpha$-attractors'. %,  to distinguish them from `polynomial   $\a$-attractors', the models to be described  in this paper, which have  potentials reaching the plateau more slowly, as inverse powers of the inflaton field. 

 After looking at Fig. \ref{LiteBIRD}, one may wonder whether there are any interesting models describing the right part of the area favored by Planck/BICEP/Keck. An interesting example of the models of this type is provided by the KKLTI attractors. Some of these   described in  \cite{Kallosh:2018zsi,Kallosh:2019hzo}  have interpretation in terms of Dp-brane inflation \cite{Dvali:1998pa,Dvali:2001fw, Burgess:2001fx,Kachru:2003sx,Lorenz:2007ze,Martin:2013tda,Kallosh:2018zsi}, though there are other ways to obtain and interpret similar potentials \cite{Stewart:1994pt,Fairbairn:2003yx,Dong:2010in,Dimopoulos:2016zhy}. 
 In particular, a broad family of such potentials have an interpretation in terms of pole inflation  \cite{Galante:2014ifa, Terada:2016nqg, Karamitsos:2019vor, Kallosh:2019hzo,Kallosh:2021mnu}, where these potentials appear as attractors. Such attractors have  potentials reaching the plateau more slowly, not exponentially but as inverse powers of the inflaton field, 
 \be \label{pol}
 V \sim V_{0}(1 -{\mu^{k}\over \vp^{k}}+... )
 \ee
where $k$ can be any (integer or not) positive constant. 

We will call these models `polynomial attractors.'  Importantly, cosmological predictions of such models in the large field limit do not depend on the detailed structure of the potential. In particular, the spectral index $n_{s}$ depends only on $k$  \cite{Kallosh:2018zsi}:
\be\label{genns}
n_{s} = 1-{2\over N_{e}}{k+1\over k+2} \ .
 \ .\ee
 Thus, investigation of any model with a potential having this behavior at large $\vp$ gives predictions that are valid for a broad class of the models of this type. That is why they are called attractors. Note that for any $k >0$ the value of $n_{s}$ is greater than the universal prediction of the exponential $\alpha$-attractors $n_{s}=   1-{2/N_{e}} \sim 0.964$ for $N_{e}\sim 55$.  In the small $k$ limit, the value of $n_{s}$ for these models can reach $1-{1/N_{e}} \sim 0.98$. For small $\mu$, these models can describe any small values of $r$, all the way down to $r = 0$. That is why the predictions of this class of  inflationary models completely cover  the right-hand side of the sweet spot of the Planck/BICEP/Keck  data  \cite{Kallosh:2019hzo,Kallosh:2018zsi,Kallosh:2019eeu}.

\begin{figure}[H]
\centering
\includegraphics[scale=0.31]{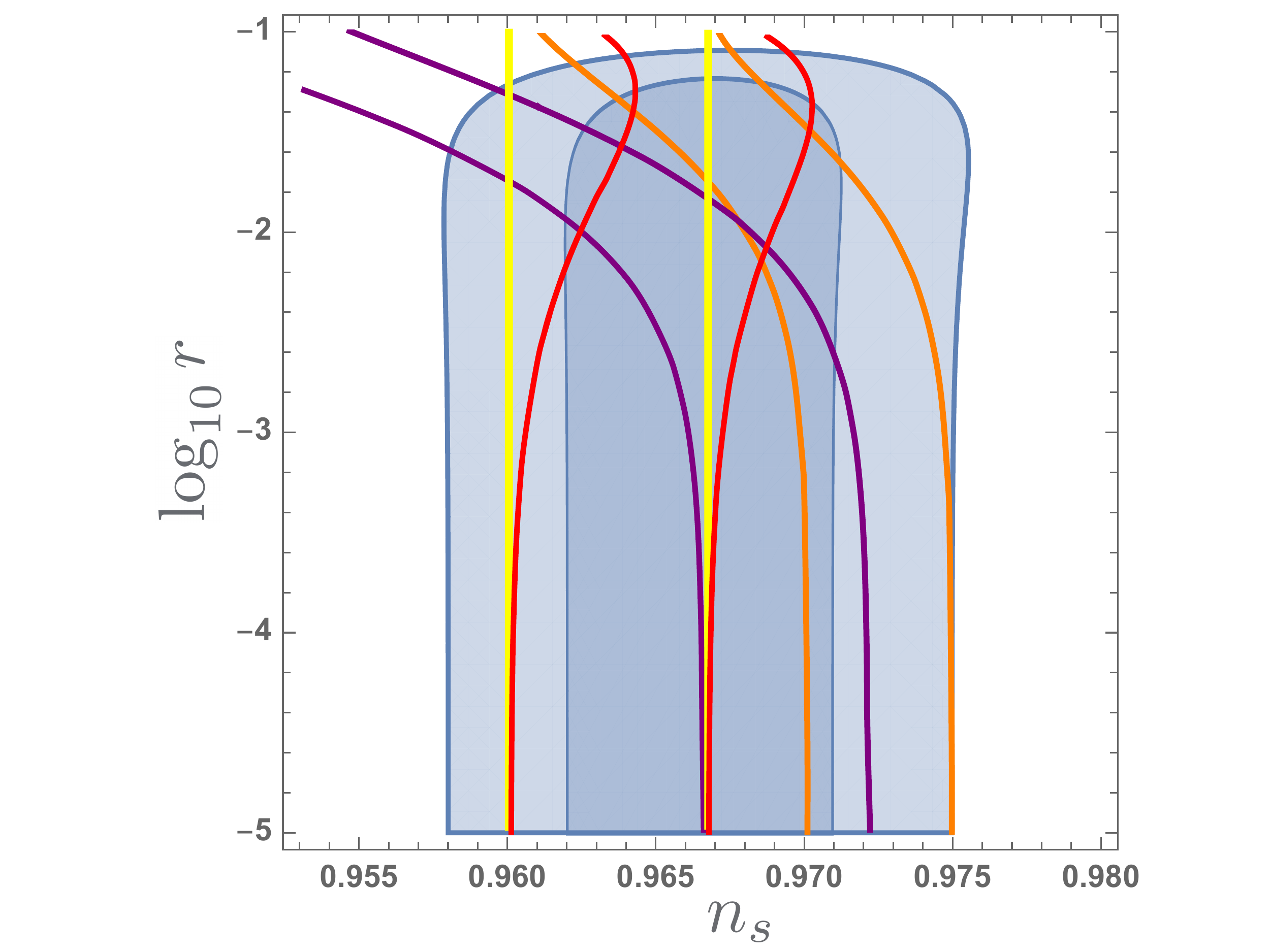}
\caption{\footnotesize 
Predictions of the simplest exponential $\alpha$-attractors and KKLTI models superimposed with the Planck2018 constraints on $n_{s}$ and $r$.  The band between the two yellow lines  is described by the simplest T-model, the two red lines correspond to the simplest E-model. 
The purple lines correspond to the quartic KKLTI model, and the orange lines describe  the quadratic KKLTI model. All bands shown here correspond to $N_{e}$ from 50 to 60.} 
\label{Blue}
\end{figure}

This is illustrated  by Fig. \ref{Blue}. The band between the two yellow lines corresponding to $N_{e}  = 50$ and $60$ is described by the simplest T-model with $V \sim \tanh^{2} {\vp\over 6\alpha}$, the two red lines correspond to E-models with $V \sim (1-e^{-\sqrt {2\over 3 \alpha}\vp})^{2}$, the purple lines correspond to the KKLTI model with  $V \sim {\vp^{4}\over  \vp^{4}+\m^{4}}$, and the orange lines describe  the model with  $V \sim {\vp^{2}\over  \vp^{2} +\m^{2}}$.

As explained in   \cite{Kallosh:2018zsi,Kallosh:2021mnu}, all of these models belong to the general class of pole inflation \cite{Galante:2014ifa,Kallosh:2018zsi}. The E- and T- models of $\a$-attractors have a pole in the kinetic term of order $q=2$, which is the consequence of the $SL(2, \mathbb{R})$  or $ SU(1,1)$ symmetry  of the kinetic terms. They have clear geometric origin corresponding to the  metric, respectively,  
\be ds^2=  -3\alpha {d T d \bar T \over (T+\bar T)^2} \ , \qquad ds^2=  -3\alpha {d Z d \bar Z \over (1-Z\bar Z)^2}\, , \qquad T = {1+Z\over 1-Z} \ .
\label{sym}\ee
Such kinetic terms often appear  in supergravity and string theory. 

In comparison, the quartic KKLTI attractors have a pole of order $q=5/2$, and the quadratic ones have $q=3$. Unlike $\alpha$-attractors, pole inflation models with $q \not = 2$ are not associated with any known symmetry and do not  originate from supergravity, though the corresponding inflaton potentials can be embedded in supergravity \cite{Kallosh:2018zsi}.

In this paper, we will make yet another step towards unification of all attractors. We will show that it is possible to incorporate the KKLTI models with $V \sim {\vp^{k}\over \vp^{k} + \m^{k}}$ in the context of $\alpha$-attractors in the framework of hyperbolic geometry based on $SL(2, \mathbb{R})$ symmetry  or $ SU(1,1)$ symmetry  of the kinetic terms \rf{sym}. In order to do it, it was necessary to consider models where the potentials of the $\alpha$-attractors had a singular first derivative at the boundary of the hyperbolic space. 

We show that the same potentials in  canonical variables are not singular. They have plateau potentials reaching the plateau as inverse powers of the inflaton field, as shown in \rf{pol}. The cosmological predictions of these models are stable with respect to the significant changes of the original potentials. In particular, their attractor predictions for $n_{s}$ are given by \rf{genns}.
 We will also present a unified supergravity model  for the exponential and polynomial $\alpha$-attractors.

 This means that now the models of such type appear  in three independent contexts: as D-brane models, as models of pole inflation with $q > 2$, and, finally, as a new family of $\alpha$-attractors. Therefore we believe that these models provide very interesting targets for the future B-mode searches. A combination of the more traditional exponential $\alpha$-attractors \cite{Kallosh:2013hoa,Ferrara:2013rsa,Kallosh:2013yoa} with the polynomial $\alpha$ attractors  to be constructed in this paper  completely covers the blue area favored by the Plank/BICEP/Keck data shown in Fig. \ref{LiteBIRD}.

\section{{  Exponential and polynomial \boldmath $\alpha$-attractors }}
\subsection{Half-plane variables}

There are many different ways to introduce $\alpha$-attractors. In the context of this paper, it is useful to start with  the pole interpretation of E-models \cite{Galante:2014ifa} assuming that the axions of the hyperbolic geometry are fixed and we can study a one-field model
\be
{\cal L} =  -{ 3\alpha\over 4} {(\partial \rho)^2\over   \rho^{2}}  - V(\rho) \ .
\label{action}\ee
As an example, one may consider a simplest potential which is  positive everywhere  and regular at  $\rho = 0$,  
\be 
V = V_0 (1- b  \rho  + \ldots)^{2}  , \qquad b > 0 \ .
\label{V}\ee
In this expression $\ldots$ represent  higher terms in $\rho$. Now we represent this theory in terms of the canonical field $\vp$, which is related to $\rho$ as follows:  
\be
\rho  = \rho_{0}\, e^{-\sqrt {2\over 3\alpha}\varphi} . 
\ee
Note that both constants $b$ and $\rho_{0}$ can be absorbed into a redefinition (shift) of the field $\vp$. Therefore without any loss of generality, the  theory \rf{action}   can be represented as
\be
{\cal L} =   - {1\over 2}  (\partial \varphi)^2- V_0 \Bigl(1 -  e^{-\sqrt {2\over 3\alpha}\varphi}+ \ldots\Bigr)^{2} . 
\ee
We called these models E-models, because of the exponential change of variables $\rho = e^{-\sqrt {2\over 3\alpha}\varphi}$. The main stage of inflation occurs at large positive values of the canonically normalized field $\varphi \gg \sqrt  \alpha$, where the omitted higher order terms are exponentially suppressed, and the potential reduces to  
\be 
V = V_0 \bigl( 1 -    e^{-\sqrt {2\over 3\alpha}\varphi}  \bigr)^{2} \ .
\label{Vcan}\ee
For $\alpha = 1$,  this potential coincides with the potential of the Starobinsky model.  
 For small $\alpha$, the cosmological predictions of $\alpha$-attractors are  
\be
\label{pred}
 n_{s} = 1-{2\over N_{e}} \ , \qquad r = {12\alpha\over N^{2}_{e}} \ .
\ee 
Note that these two results, in the large $N_{e}$ (or small $\alpha$) limit, do not depend on the detailed structure of $V(\rho)$. Our main assumption was that the potential is a relatively simple function that  can be expanded  in Taylor series at $\rho = 0$.

 Once  this restriction is removed, the situation may change. For example, if  potential $V(\rho)$ is singular at the boundary, the inflaton potential may change dramatically, unless the corresponding corrections are strongly suppressed. In general, it may be possible to use these models constructively. In particular, such models may simplify the solution of the problem of initial conditions for inflation \cite{Linde:2017pwt,Linde:2018hmx}. Other applications of such models have been considered for example in  \cite{Garcia-Saenz:2018ifx,Iacconi:2021ltm}. We will not study  models with a singular potential here, but instead we will make one relatively small step: We will consider models where {\it the potential is non-singular, but its derivative is singular}. This means that $b = V' (\rho)\rightarrow \infty$ near the boundary $\rho\rightarrow 0$. Such modifications preserve the plateau shape of the potential, but may have other important consequences.

As a simplest example, let us add a term $\sim \sqrt \rho$ to the potential $V(\rho)$ \rf{V}:
\be 
V = V_0 (1- a\sqrt{\rho} - b \rho + \ldots)^{2} \ .
\label{V2}\ee
This potential is non-singular at small $\rho$, but its derivative is singular. In canonical variables we have
\be 
V = V_0 (1- ae^{- \sqrt{1\over 3\alpha}\varphi} - b e^{-\sqrt {2\over 3\alpha}\varphi} + \ldots)^{2}   .
\label{V2a}\ee
At large $\vp$ (after the shift of $\vp$ absorbing $a$),  we have 
\be 
V = V_0 (1-  e^{- \sqrt{1\over 3\alpha}\varphi})^{2}   \ .
\label{V2b}\ee
Thus once again we have the E-model similar to \rf{Vcan}, but its effective $\alpha$ doubled.  This case shows that the removal of restriction of Taylor series at $\rho = 0$ changes the predictions. In this particular case, $n_s$ remains  the same, we still have a model in a class of the exponential $\alpha$-attractors,  but the value of  $r$ will be $24\alpha/ N^{2}_{e} $,  two times greater than in  \rf{pred}.

Note, that this model is still a cosmological attractor. In fact, it is a stronger attractor than the model \rf{V}, because the predictions of the new model will not change not only if we add there any terms $O(\rho^{2})$, but also if we add to it any terms $O(\rho)$, which are much greater than $O(\rho^{2})$ in the important limit $\rho \to 0$ corresponding to $\vp \to \infty$.

Now we will make the next step and consider  the logarithmic potential  
\be 
V = V_0\,   {\ln^2\rho \over  \ln^2\rho+ c^{2 }}   \ .  
\label{Vlog}\ee
As before, this  potential is finite at $\rho \to 0$, even though its derivative diverges there. In canonical variables, it is given by
\be 
V = V_0\,   { \varphi^{2} \over \varphi^{2}+ \m^{2}}  \ .  
\label{Vcan0}\ee
where 
\be
\mu = \sqrt{3\alpha\,\over 2} c  \ .
\label{mu}\ee
This potential has a plateau of height $V_{0}$. It is  a legitimate $\alpha$-attractor,  in more ways than one, although it is different from 
the exponential $\alpha$-attractors.  First of all, the behavior of \rf{Vcan0}  in the large $\vp$ limit does not change  if we add any terms $\rho^{\nu}$ with $\nu > 0$ to the potential \rf{Vlog},  or to the denominator or nominator   in \rf{Vlog}.  Secondly, the parameter $\m^{2}$ vanishes in the limit $\alpha \to 0$. The potential at large $\vp$ approaches the plateau as 
\be
V = V_0\Big (1- {\mu^2\over \vp^2}+\dots\Big) = V_0\Big (1- {3\alpha c^2\over 2 \vp^2}+\dots\Big)\ .
\ee
Thus, in the new attractors the plateau is described by  inverse powers of $\vp$. For brevity, we called these attractors polynomial  to distinguish them from the previously known attractors, where the approach to a plateau was exponentially fast \rf{V2b}.

At small $\m$, the parameters $n_{s}$ and $r$ reach their attractor values \cite{Martin:2013tda,Kallosh:2018zsi,Kallosh:2019hzo}
\be 
n_{s} =  1-{3\over 2N_{e}} ,\qquad r  = {\sqrt 2 \m \over N_{e}^{3/2}} =  { \sqrt {3\alpha}\, c \over N_{e}^{3/2}} \ . 
\label{nsr1}\ee

The next case to be considered here is 
\be 
V = V_0\,  {\ln^4\rho \over  \ln^4\rho+ c^{4 }} \ .  
\label{Vlog4}\ee
The  potential is finite at $\vp \to 0$, even though its derivative diverges there. In canonical variables, this potential is given by
\be 
V = V_0\,   { \varphi^{4} \over \varphi^{4}+ \m^{4}}\, , 
\label{Vcan4}\ee
where
$ 
\mu = \sqrt{3\alpha\,\over 2} c.
$
This potential has a plateau of height $V_{0}$. The potential at large $\vp$ approaches the plateau as $V_0\Big (1- {\mu^4\over \vp^4}+\dots\Big)$.
At small $\alpha$, the parameters $n_{s}$ and $r$ reach their attractor values \cite{Kachru:2003sx,Martin:2013tda,Kallosh:2018zsi,Kallosh:2019hzo}
\be
n_{s} =  1-{5\over 3N_{e}} , \qquad    r  = {  4 \m^{4/3} \over (3N_{e})^{5/3}}  = {  (4\alpha c^{2})^{{2/3}}   \over 3   N_{e}^{5/3}}  \ .
\label{nsr2}\ee  
Similarly, one may consider potentials 
\be 
V = V_0\,  {\ln^{2n}\rho \over  \ln^{2n}\rho+ c^{2n }} \ .  
\label{Vlog4a}\ee
In canonical variables, this potential is given by
\be 
V = V_0\,   { \varphi^{2n} \over \varphi^{2n}+ \m^{2n}}\, . 
\label{Vcan4a}\ee
One may also consider another class of potentials,
\be 
V = V_0\,   {( \ln^2\rho +c^{2})^{k/2}- c^{k}\over  ( \ln^2\rho +c^{2})^{k/2}+ c^{k}}   \ ,  
\label{Vlonglog}\ee
where $k$ and $c$ are any positive numbers. In canonical variables, this potential becomes
\be 
V = V_0\,   {(\vp^{2} +\mu^{2})^{k/2}- \mu^{k}\over  (\vp^{2} +\mu^{2})^{k/2}+ \mu^{k}}   \ ,  
\label{Vlongcan}\ee
where, as before, $\mu = \sqrt{3\alpha\,\over 2} c $. At large $\vp$, the potential and its predictions for $n_{s}$ are given by \rf{pol}, \rf{genns}:
\be \label{pol2}
 V \sim V_{0}(1 -{\mu^{k}\over \vp^{k}}+... ) \ , \qquad n_{s} = 1-{2\over N_{e}}{k+1\over k+2} \ .
 \ee
Note that for any $k> 0$ the attractor values of $n_{s}$ in \rf{pol2} are greater than $1-2/N_{e}$, and in the limit $k \to 0$ one has $n_{s} \to  1-1/N_{e}$.

\subsection{Disk variables}
Note that the kinetic term in eq. \rf {action} originates from the hyperbolic geometry in half-plane variables
\be
 ds^2=  -3\alpha {d T d \bar T \over (T+\bar T)^2} \ .
 \ee
In Poincar\'e disk variables the geometry is 
\be
 ds^2= -3\alpha {dZ d\bar Z\over (1-Z\bar Z)^2} \ .
  \ee
The Cayley transform from half-plane to disk variables, in our approximation that the axions in hyperbolic geometry are stabilized, can be taken in the form
\be
\rho = {1+z\over 1-z} \ ,
\ee
where $z=\tanh
 {\varphi \over \sqrt{6\alpha} } 
$.
In such case the potentials for polynomial  $\alpha$-attractors
in disk coordinates are
\be 
V = V_0\,  {\ln^{2n}{1+z\over 1-z} \over  \ln^{2n}{1+z\over 1-z}+ c^{2n }}     \ . 
\label{VlogDisk}\ee
The canonical potentials are the same as in equation   \rf{Vcan4a}.  Similarly, one can use disk variables to describe the broad class of potentials of the type of \rf{Vlongcan}, \rf{pol2}.

\section{Supergravity version of exponential and polynomial {\boldmath $\a$-attractors}} 
In the previous section, we outlined the embedding of exponential and polynomial $\a$-attractors in hyperbolic geometry. Here we  present  the supergravity version of both types of models using the construction proposed in \cite{Kallosh:2017wnt}, which we called the Model Building Paradise.  It was further developed in \cite{Gunaydin:2020ric,Kallosh:2021vcf}. %and for the case of one inflaton multiplet ($T$ for half-plane case and $Z$ for disk case) and a nilpotent one $X$ with $X^2=0$ it is explained in  \cite{Kallosh:2021fvz}.
A concise description of the method with various illustrative examples for the case of one inflaton multiplet ($T$ for half-plane case and $Z$ for disk case) and a nilpotent one $X$ with $X^2=0$  can be found in \cite{Kallosh:2021fvz}.

\

For models in {\it half-plane variables} $T$ with stabilized axions one can take the following \K potential and superpotential
\be
K(T,\overline T)= - 3\a \log  (T + \overline{   T})+   {F_X^2 \over F_X^2 +   V_{\rm infl}(T,\overline T) } X\overline X \ , \quad 
W  =  (W_0+ F_X X)(2T)^{3\alpha/2} \ ,
\label{KW}\ee
which yields
\be\label{infpot3}
V_{\rm total} (Z) = \Lambda + V_{\rm infl}(T,\overline T)_{|_{T=\overline T = t}} \, , \qquad \Lambda=F_X^2 -3W_0^2\, , \qquad t= {T+\overline T\over 2}= e^{-\sqrt{2\over 3\alpha}\vp} \ .
\ee
For
$
V_{\rm infl}(T, \bar T)  = m^{2}(1-T)(1-\bar T)
$ we find  the simplest E-model of $\a$-attractors with the  potential 
$
V_{\rm total}(\vp) =\Lambda +  m^{2}\bigl(1- e^{-\sqrt{2\over 3\alpha}\phi}\bigr)^{2}
$.

For $V_{\rm infl}(T,\overline T)= V_0 { \ln^{2n}  t \over c^{2n}+\ln^{2n}  t}$
one finds  a family of polynomial   $\a$-attractors with the potential   $V(\vp) = \Lambda +  V_0 {\vp^{2n}\over \m^{2n} + \vp^{2n}}
$. This is the same potential as in \rf{Vcan4a}, but now it   includes an arbitrary cosmological constant $\Lambda$.

\

For models in {\it disk variables} $Z$  we take 
\be\label{KT}
K(Z,\overline Z) =- 3\alpha \log  (1 - Z\overline{Z})  +   {F_X^2 \over F_X^2 +   V_{\rm infl}(Z,\overline Z) } X\overline X, \quad  W=  (W_0+ F_X X) (1-Z^{2})^{3\alpha/2} \ ,
\ee
which yields
\be\label{infpot3a}
V_{\rm total} (Z) = \Lambda + V_{\rm infl}(Z,\overline Z)_{|_{Z=\overline Z = z}} \, , \qquad \Lambda=F_X^2 -3W_0^2\, , \qquad z^2= Z \overline Z= \tanh^2{\vp \over \sqrt{6\alpha}} \ .
\ee
For   $V_{\rm infl}(Z,\overline Z)= m^2 z^2$, 
this leads to the simplest T-models of  exponential $\a$-attractors with  $V(\vp) = \Lambda + m^2 \tanh^{2}{\vp\over \sqrt{6\alpha}}$.

For $V_{\rm infl}(Z,\overline Z)_{|_{Z=\overline Z = z}}= V_0 { \ln^{2n}  [{1+z\over 1-z}]\over c^{2n}+\ln^{2n}  [{1+z\over 1-z}]}$
this leads, once again, to the family of polynomial   $\a$-attractors with the potential  
$V(\vp) = \Lambda +  V_0 {\vp^{2n}\over \m^{2n} + \vp^{2n}}.
$

\section{Discussion} 

As we already mentioned in the Introduction, there are two main types of inflationary models with plateau potentials. The potentials which appear in the Starobinsky model, GL model, Higgs inflation, and in T- and E- models of $\alpha$-attractors have the same basic structure at large positive $\vp$ shown in \rf{exp}: Their deviation from the plateau decreases exponentially fast at   large $\vp$.  We called such models exponential attractors. 

These models are well known and well explored. Their predictions are stable with respect to significant modifications of the  models. In particular, all of these models, independently of their physical origin and interpretation, have the same attractor prediction $n_{s} = 1-2/N_{s}$ in the large $N_{e}$ limit, consistent with the Planck/BICEP/Keck results.  In addition, T- and E- models of $\alpha$ attractors can describe any small values of $r$, and can be formulated in the theories with hyperbolic geometry, which is often encountered in supergravity and string theory. Advanced versions of $\alpha$ attractors have 7 different targets for $r$ in the most interesting range $10^{-3} < r < 10^{-2}$.   

%Predictions of these models fully cover the left hand side of the blue area in 

The second class of attractors with plateau potentials have the potentials reaching the plateau more slowly, like  $V_{0}(1 -{\mu^{k}/\vp^{k}})$  \rf{pol}. We  called these models `polynomial attractors.'  Cosmological predictions of such models in the large field limit also do not depend on the detailed structure of the potential. In particular, the spectral index $n_{s}$ depends only on $k$  \cite{Kallosh:2018zsi}.
 
 Some of these models are called KKLTI; they may be related to Dp-brane inflation \cite{Dvali:1998pa,Kachru:2003sx,Martin:2013tda}. A   broad class of such models can be incorporated in the general theory of pole inflation with the pole of the kinetic term of degree $q > 2$ \cite{Galante:2014ifa, Kallosh:2018zsi}. However, until now we did not know whether it is possible to develop these models even further and make them a part of the broad family of $\a$-attractors. 
 
 In this paper we analyzed this issue and found a large class of polynomial $\alpha$-attractors. The technical reason why they are different from the exponential $\alpha$-attractors has to do with the properties of the potentials and their derivatives near the boundary of the Poincar\'e disk, as explained in Sec. 2. 
 
 These potentials
  include, in particular, the potentials of the type of  ${\vp^{2n} \over \vp^{2n}+ \mu^{2n}}$ with $\mu^2 = {3\alpha\over 2}c^{2}$.  As a result, now such models have  three different, independent interpretations. They appear in the context of Dp-brane inflation, and in the context of pole inflation, and now they also belong to a special class of $\alpha$-attractors.  Therefore we believe that these models provide very interesting targets for the future B-mode searches.
 
\begin{figure}[H]
\centering
\includegraphics[scale=0.1]{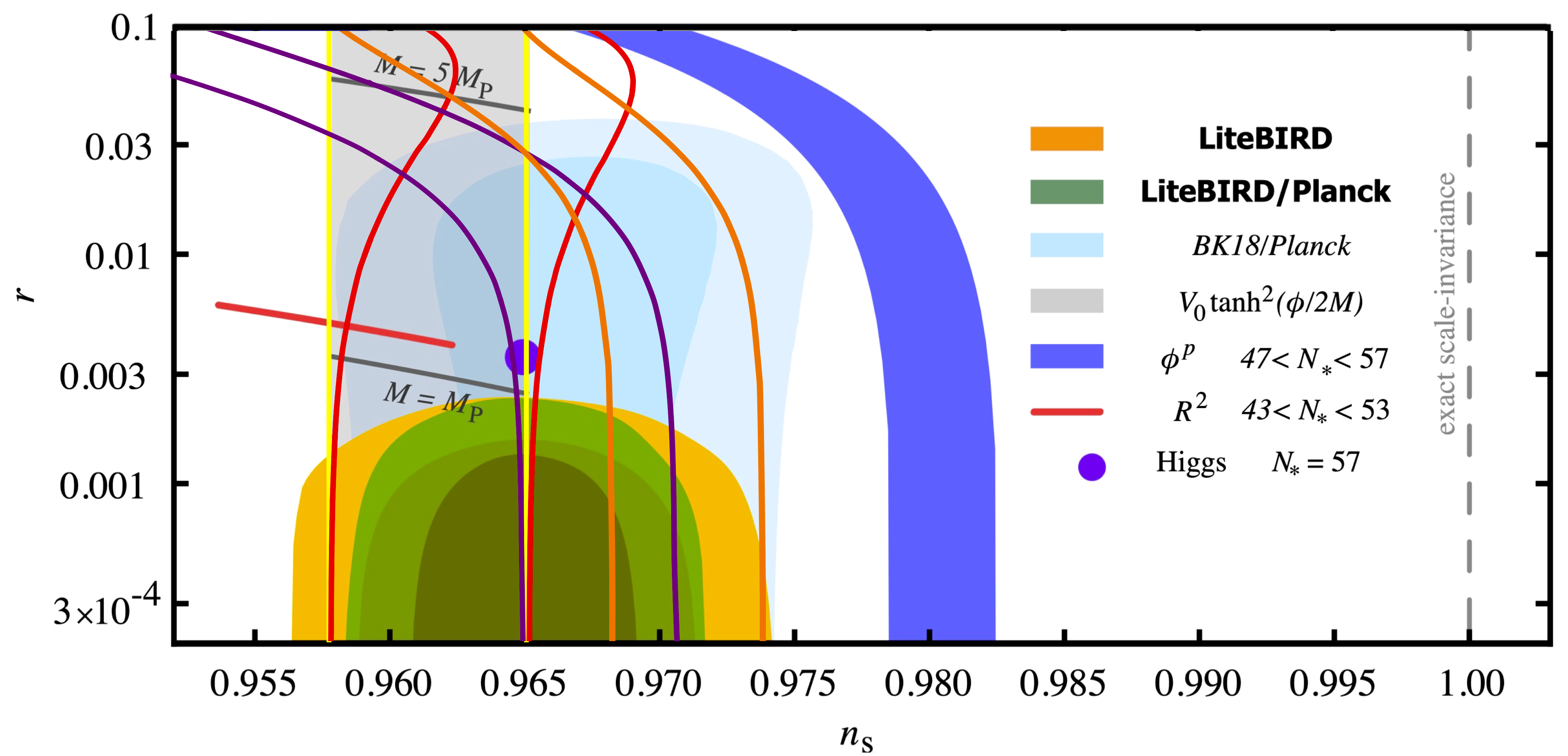}
\caption{\footnotesize We added predictions of the two simplest polynomial $\alpha$-attractors \rf{Vcan0}, and \rf{Vcan4} to the   LiteBIRD figure Fig. \ref{LiteBIRD}.  As before, the gray area shows the predictions of the simplest $\alpha$-attractor T-models with the potential \rf{T}. It is surrounded by two yellow lines corresponding to the number of e-foldings $N_{e} = 47, 57$. The two   red lines for $N_{e} = 47, 57$ surround the predictions of  E-models of $\alpha$-attractors with the potential \rf{E}. The purple and orange lines represent  the polynomial $\a$-attractors ${\vp^{4} \over \vp^{4}+ \mu^{4}}$ \rf{Vcan4}, and  ${\vp^{2} \over \vp^{2}+ \mu^{2}}$ \rf{Vcan0}, for $N_{e} = 47$ and 57.
}
 \label{LiteBIRDnew}
\end{figure}

 To explain the phenomenological implications of these results, we  added the  two simplest polynomial $\alpha$-attractor  models \rf{Vcan0}    and \rf{Vcan4} to the  LiteBIRD figure Fig. \ref{LiteBIRD}  shown in the beginning of this paper. The results are shown in Fig. \ref{LiteBIRDnew}.
%The two additional attractors between the purple and orange lines are polynomial $\a$-attractors.  
As one can see from Fig. \ref{LiteBIRDnew}, the two simplest T- and E- models \rf{T} and \rf{E}   in combination with the two simplest polynomial $\alpha$-attractors \rf{Vcan0}, and \rf{Vcan4} completely cover the dark blue area favored by the   latest Planck/BICEP/Keck results.

 \section*{Acknowledgement}
We are grateful to   S. Ferrara,  F. Finelli, R. Flauger,  C. L. Kuo, C. Pryke, D. Roest, T. Wrase and Y. Yamada  for many stimulating discussions.  This work is  supported by SITP and by the US National Science Foundation Grant  PHY-2014215.  

\newpage

\bibliographystyle{JHEP}
\bibliography{lindekalloshrefs}
\end{document}